\documentclass[aps,epsfig,prd,twocolumn,superscriptaddress,nofootinbib]{revtex4-1}
\usepackage{graphicx}
\usepackage{natbib}
\usepackage{color}
\usepackage{array}
\usepackage{amssymb}

\begin{document}

\title{New constraints on $f(R)$ gravity from clusters of galaxies}

\author{Matteo Cataneo}\email{matteoc@dark-cosmology.dk}
\affiliation{Dark Cosmology Centre, Niels Bohr Institute, University of Copenhagen, Juliane Maries Vej 30, 2100 Copenhagen, Denmark}
\affiliation{SLAC National Accelerator Laboratory, 2575 Sand Hill Road, Menlo Park, CA 94025, USA}
\affiliation{Kavli Institute for Particle Astrophysics and Cosmology, Stanford University, 452 Lomita Mall, Stanford, CA 94305, USA}
\affiliation{Stanford Institute for Theoretical Physics, Stanford University, 382 Via Pueblo Mall, Stanford, CA 94305, USA}

\author{David Rapetti}
\affiliation{Dark Cosmology Centre, Niels Bohr Institute, University of Copenhagen, Juliane Maries Vej 30, 2100 Copenhagen, Denmark}

\author{Fabian Schmidt}
\affiliation{Max-Planck-Institut f\"ur Astrophysik, Karl-Schwarzschild-Str. 1, D-85741 Garching, Germany}

\author{Adam B. Mantz}
\affiliation{Department of Astronomy and Astrophysics, University of Chicago, 5640 South Ellis Avenue, Chicago, IL 60637, USA}
\affiliation{Kavli Institute for Cosmological Physics, University of Chicago, 5640 South Ellis Avenue, Chicago, IL 60637, USA}

\author{Steven W. Allen}
\affiliation{SLAC National Accelerator Laboratory, 2575 Sand Hill Road, Menlo Park, CA 94025, USA}
\affiliation{Kavli Institute for Particle Astrophysics and Cosmology, Stanford University, 452 Lomita Mall, Stanford, CA 94305, USA}
\affiliation{Department of Physics, Stanford University, 382 Via Pueblo Mall, Stanford, CA 94305, USA}

\author{Douglas E. Applegate}
\affiliation{Argelander-Institute for Astronomy, Auf dem H\"ugel 71, D-53121 Bonn, Germany}

\author{Patrick L. Kelly}
\affiliation{Department of Astronomy, University of California, Berkeley, CA 94720, USA}

\author{Anja von der Linden}
\affiliation{Dark Cosmology Centre, Niels Bohr Institute, University of Copenhagen, Juliane Maries Vej 30, 2100 Copenhagen, Denmark}
\affiliation{Kavli Institute for Particle Astrophysics and Cosmology, Stanford University, 452 Lomita Mall, Stanford, CA 94305, USA}
\affiliation{Department of Physics, Stanford University, 382 Via Pueblo Mall, Stanford, CA 94305, USA}

\author{R. Glenn Morris}
\affiliation{SLAC National Accelerator Laboratory, 2575 Sand Hill Road, Menlo Park, CA  94025, USA}
\affiliation{Kavli Institute for Particle Astrophysics and Cosmology, Stanford University, 452 Lomita Mall, Stanford, CA 94305, USA}

\date{\today}

\begin{abstract}
The abundance of massive galaxy clusters is a powerful probe of departures from General Relativity (GR) on cosmic scales. Despite current stringent constraints placed by stellar and galactic tests, on larger scales alternative theories of gravity such as $f(R)$ can still work as effective theories. Here we present constraints on two popular models of $f(R)$, Hu-Sawicki and ``designer", derived from a fully self-consistent analysis of current samples of X-ray selected clusters and accounting for all the covariances between cosmological and astrophysical parameters. Using cluster number counts in combination with recent data from the cosmic microwave background (CMB) and the CMB lensing potential generated by large scale structures, as well as with other cosmological constraints on the background expansion history and its mean matter density, we obtain the upper bounds $\log_{10}|f_{R0}| < 4.79$ and $\log_{10}B_0 < 3.75$ at the 95.4 per cent confidence level, for the Hu-Sawicki (with $n=1$) and designer models, respectively. The robustness of our results derives from high quality cluster growth data for the most massive clusters known out to redshifts $z \sim 0.5$, a tight control of systematic uncertainties including an accurate and precise mass calibration from weak gravitational lensing data, and the use of the full shape of the halo mass function over the mass range of our data.
\end{abstract}


\maketitle

\section{Introduction}

Since the discovery of the late time cosmic acceleration
\cite{1999ApJ...517..565P,1998AJ....116.1009R} a profusion of
theoretical models have been proposed to explain this phenomenon (for recent reviews see
\cite{2006IJMPD..15.1753C,2012PhR...513....1C,2015PhR...568....1J}). In
a nutshell, one can either add a dark fluid with sufficient
negative pressure or modify the laws of gravity. Among the alternative
theories to General Relativity (GR), $f(R)$ gravity has sparked a lot of
interest over the last decade, motivated by its relative simplicity and
rich phenomenology
\cite{2010RvMP...82..451S,2010LRR....13....3D}. In this model, the
Einstein-Hilbert action is supplemented by a non-linear function of
the Ricci or curvature scalar, $R$. Conveniently chosen $f(R)$
functions can reproduce the observed accelerated expansion while adding an
attractive force of the order of the gravitational interaction. This fifth force is carried by the scalar degree of freedom, dubbed
scalaron, $f_R = \text{d}f/\text{d}R$, introduced by the modification
of gravity. The range of this new interaction is given by the inverse
mass, or equivalently the Compton wavelength of the scalaron, which is
directly related to the background amplitude of the scalaron field today, $f_{R0}$.\\
\indent In this model, on scales smaller than the Compton wavelength,
gravity is enhanced by a factor of 4/3 and structure formation is consequently
modified. Above this scale, structures assemble following GR as long as the background Compton wavelength is smaller than the horizon, $\lambda_C \ll H^{-1}$. \\
\indent Viable $f(R)$ models also present a non-linear
mechanism to suppress the modifications of gravity in high-density environments,
such as in our Solar System, where GR is known to be a very accurate
theory of gravity. This suppression should also be
observed within our Galaxy. Theoretical arguments \cite{2007PhRvD..76f4004H} supported afterwards by hydrodynamical simulations of galaxy formation and evolution \cite{2013MNRAS.436.2672F} require the
value of the background field $|f_{R0}|$ to be less than $10^{-6}$ for this to be the case. Most recently, constraints from distance
indicators and dwarf galaxies further reduced this upper limit 
to $|f_{R0}| \lesssim 4\times 10^{-7}$ (here and throughout, we state the upper limits at the 95.4 per cent confidence level) \cite{2013ApJ...779...39J,2013JCAP...08..020V}.  Such small  $f(R)$
modifications of gravity  cannot leave their imprints on cosmological
scales or even on fully non-linear scales such as those within galaxy
clusters. Nevertheless, $f(R)$ can serve as a useful effective theory or working model for tests of gravity on
large scales. For this purpose, clusters of galaxies represent a powerful probe of gravity down to scales $\sim 1$--$20$ Mpc/$h$. In particular, it has been shown \cite{2009PhRvD..79h3518S,2014JCAP...03..021L} that the abundance of rare massive halos is substantially enhanced by the presence of a fifth force for $|f_{R0}| > |\Psi| \sim 10^{-6}$--$10^{-5}$, where $\Psi$ is the typical depth of the Newtonian potential for these objects.\\
\indent In combination with other data sets, \citet{2009PhRvD..80h3505S} used measurements of the abundance of massive galaxy clusters inferred from X-ray survey data to constrain the Hu-Sawicki model of $f(R)$ gravity \cite{2007PhRvD..76f4004H} and obtained the tightest cosmological constraint at the time $|f_{R0}|\lesssim1.3\times 10^{-4}$. These authors used a spherical collapse prediction of the number of halos as a function of cosmological parameters, mass and redshift that had previously been validated using N-body simulations \cite{2009PhRvD..79h3518S}. We employ this halo mass function (HMF) and extend the approach by including departures from GR as a prefactor to the HMF of \citet{2008ApJ...688..709T}, which is based on high resolution GR simulations. This method allows us to efficiently use the full HMF of GR as a baseline, properly accounting for the redshift evolution and covariances of its parameters, as well as other systematic uncertainties (see e.g.~\cite{2010MNRAS.406.1759M}). In~\cite{2009PhRvD..80h3505S}, the authors mapped modifications of gravity into GR by matching the Sheth-Tormen (ST) HMF \cite{1999MNRAS.308..119S} for $f(R)$ to a Tinker \textit{et al.}\ mass function with rescaled $\sigma_8$ at a fixed pivot mass. This renormalization was then used to incorporate both CMB and cluster constraints on the growth of structures. These simplifications allowed them to have a limited number of parameters and therefore to be able to perform a maximum likelihood analysis. However, this approach may neglect relevant correlations between astrophysical and cosmological quantities as well as introduce spurious degeneracies between them. Here instead we carry out a Markov Chain Monte Carlo (MCMC) analysis of the full likelihoods of current cluster and CMB data sets, which includes all the covariances between parameters and an advanced treatment of systematic uncertainties and biases. Together with CMB data, and using the full mass and redshift dependence of the HMF, as well as high quality survey (X-ray) and extensive follow-up (X-ray and optical) cluster data, spanning a redshift range $0 < z < 0.5$, we obtain robust and improved constraints on the background scalaron field, $|f_{R0}|<1.6\times 10^{-5}$. As in \cite{2009PhRvD..80h3505S}, our results also include constraints from baryon acoustic oscillation (BAO) and type Ia supernova (SNIa) data. \\
\indent More recently, \citet{2014JCAP...03..046D} and \citet{PhysRevD.91.063524} obtained somewhat tighter upper bounds on $|f_{R0}|$ by comparing the theoretical predictions of the enhanced linear matter power spectrum in $f(R)$ gravity with measurements of the galaxy power spectrum made by the WiggleZ Dark Energy Survey \cite{2010MNRAS.401.1429D}. As described in those analyses, however, $f(R)$ corrections for the non-linear scales of the matter power spectrum (see e.g. \cite{2013MNRAS.428..743L}) and for the scale-dependence of the halo bias \cite{2011PhRvD..83f3511P} were not included. Note that in these as well as in our work, a uniform prior on the logarithm of either the background scalaron field or its Compton wavelength at the present epoch is used in obtaining the main results. We show here that a different choice of prior (e.g. uniform on $f_{R0}$) can in practice have a non negligible effect on the constraints (see section~\ref{sec:results}).\\
\indent For the ``designer'' $f(R)$ model, using data from cluster number counts and a uniform prior on the Compton wavelength in Hubble units ($B_0$), \citet{2012PhRvD..85l4038L} placed an upper limit on this parameter that is equivalent to $|f_{R0}| < 2 \times 10^{-4}$. Unlike previous works, that paper used optically selected clusters from the Sloan Digital Sky Survey (SDSS) data \cite{2007ApJ...660..239K}. Moreover, the modifications of gravity were included in the Tinker et al.\ HMF (based on GR) through only the calculation of the variance of the linear matter density field. The authors justified this approach by arguing that the data were not sufficiently constraining to enter the regime $|f_{R0}| < 10^{-4}$, where such a HMF is known to no longer be accurate enough.\\
\indent Secondary anisotropies of the CMB can also be used to measure modifications of gravity. The enhancement in the growth of structure due to $f(R)$ gravity has potentially observable effects on linear scales through the Integrated Sachs-Wolfe (ISW) effect and CMB lensing \cite{2006PhRvD..73l3504Z,2007PhRvD..76f3517S,2014JCAP...03..046D}. Recent measurements by the \textit{Planck} satellite of the CMB lensing potential generated by large scale structures\footnote{Note that these measurements are statistically independent of those from the temperature power spectrum in that the lensing potential power spectrum is a higher-order correlation function of the CMB temperature maps (see \cite{2006PhR...429....1L,2011PhRvL.107b1301D,2012ApJ...756..142V,2014A&A...571A..17P} for details).} together with CMB temperature and polarization data place a weak upper bound on $f(R)$ modifications, $|f_{R0}| < 10^{-3}$ \cite{2014PhRvD..90d3513R}. This additional power is included in our analysis, and for CMB data alone we find consistent results with previous works. Furthermore, combining CMB with cluster data helps break parameter degeneracies and tightens significantly the constraints on the normalization of the matter power spectrum, $\sigma_8$. This information is fully accounted for in our results through the multidimensional parameter covariance provided by our joint likelihood analysis.\\
\indent This paper is organized as follows. In \S \ref{sec:model} we review the phenomenology of $f(R)$ gravity and briefly describe its popular models, Hu-Sawicki \cite{2007PhRvD..76f4004H} and designer \cite{2007PhRvD..75d4004S,2008PhRvD..77b3503P}. In \S \ref{sec:hmf} we discuss the halo mass function employed here. \S \ref{sec:data} contains a description of our cluster data sets, as well as of the other cosmological data sets with which we combine them. Finally, we present our results in \S \ref{sec:results} and conclude in \S \ref{sec:conclusions}.

\section{$f(R)$ gravity \label{sec:model}}

In this work we constrain modified gravity theories for which the
Einstein-Hilbert action in the Jordan frame includes a general
non-linear function of the Ricci scalar, such as

\begin{equation}
S_{EH} = \int \, \text{d}^4 x\sqrt{-g} \left[ \frac{R + f(R)}{16 \pi G} \right].
\end{equation}

\noindent Here and throughout, we set $c = 1$.  GR
with a cosmological constant $\Lambda$ is recovered for $f = -2\Lambda$. 
This gravity model
exhibits an additional attractive force mediated by a new scalar
degree of freedom, the scalaron field $f_R \equiv \text{d}f /
\text{d}R$. For viable $f(R)$ models (see
e.g.\ \cite{2007PhRvD..76f4004H,2008PhRvD..77b3503P}), its range is
given by the physical Compton wavelength $\lambda_C = (3 \, \text{d}f_R /
\text{d}R)^{1/2}$. One of the effects of this fifth force is the
enhancement of the abundance of massive dark matter halos, as described
in \S \ref{sec:hmf}. However, such modifications of gravity are suppressed by
the non-linear chameleon effect in high density regions, where the
depth of the gravitational potential wells is large compared to the
background field, $|\Psi| > |f_R(\bar{R})|$. Note that, throughout the text, overbars denote background quantities. \\
\indent Previous analytical and numerical works \cite{2014PhRvD..89b3521N,2012PhRvD..86l3503H,2008PhRvD..78l3523O} have shown that for $|f_{R0}| \ll 1$, time derivatives of the scalar field can be neglected compared to spatial derivatives, making the quasi-static approximation (QSA) a fairly accurate description of the modified dynamics on all scales. Relaxing this approximation yields effects of the order $\lambda_C^2/H^{-2}$, which could be significant for $|f_{R0}| \sim 1$ at large scales \cite{2012PhRvD..86l3503H}. However, the ISW effect is the only known observable at (near)-horizon scales, and the authors in \cite{2012PhRvD..86l3503H} showed that it is actually insensitive to large scale corrections associated with the evolution of the scalaron field. Note also that cluster scales are well within the horizon, and hence are not affected by the QSA approximation.\\
\indent Since $f(R)$ gravity is conformally equivalent to a scalar-tensor theory with constant coupling to the matter fields, whereas electromagnetism is conformally invariant, the geodesics of photons are unchanged by this modification of gravity apart from a conformal rescaling of the gravitational constant by $1+f_R$ \cite{1994ApJ...429..480B}. In other words, given a fixed density field, e.g.\ a halo of mass $M$, the resulting lensing potential shows no deviation from that in GR as long as $|f_R| \ll 1$. This argument is particularly important for our observed mass function, since we currently employ a weak gravitational lensing analysis to calibrate our cluster masses. For the field values of interest here ($|f_R| \ll 1$), the assumption of GR in the lensing analysis is conveniently valid for our calculations.\\
\indent Each $f(R)$ model produces its own evolution of $\lambda_C$  \cite{2011PhRvD..83f3503F}, and the corresponding chameleon
screening becomes active at a different redshift and degree of
non-linearity, impacting accordingly the growth of structures
(cf. \cite{2013PhRvD..88j3507H,2013MNRAS.428..743L,2010JCAP...12..006A}). Here we consider two popular forms of $f(R)$, the Hu-Sawicki (HS)
\cite{2007PhRvD..76f4004H} and ``designer" models
\cite{2007PhRvD..75d4004S,2008PhRvD..77b3503P}. \\

\subsection {Hu-Sawicki model}

The HS models have the following functional form    

\begin{equation}
f(R) = -2\Lambda\frac{R^n}{R^n+\mu^{2n}},
\end{equation}

\noindent with $\Lambda$, $\mu^2$ and $n$ being free parameters. Note
that since $R \rightarrow 0$ implies $f(R) \rightarrow 0$ this model does not strictly contain a cosmological constant. However, in the high-curvature regime, $R \gg \mu^2$, the function above can be approximated as 

\begin{equation}\label{HSapprox}
f(R) = -2\Lambda - \frac{f_{R0}}{n}\frac{\bar{R}_0^{n+1}}{R^n}.
\end{equation}

\noindent $f_{R0} = -2n\Lambda \mu^{2n}/\bar{R}_0^{n+1}$, which replaces $\mu^2$ as a free parameter of the model, and $\bar{R}_0 \equiv \bar{R}(z=0)$, so that $f_{R0} = f_R({\bar{R}_0})$. Notice that, for $|f_{R0}| \ll 1$, the curvature scales set by $\Lambda \sim \mathcal{O}(\bar{R}_0)$ and $\mu^2$ are very different. This guarantees the validity of the $R \gg \mu^2$ approximation today and in the past. \\
\indent For this model, deviations from a cosmological constant are of
the order of $f_{R0}$. Consequently, in the limit $|f_{R0}| \ll
10^{-2}$, HS closely mimics the $\Lambda$CDM expansion history making
these two models practically indistinguishable by geo\-metric
tests. However, $f_{R0}$ also affects the formation of cosmic structures. If we fix the scaling index $n$, geometric
probes can constrain $\Lambda$, whereas growth tests, such as cluster
abundance, can constrain $f_{R0}$, which controls the strength and range of the force modification. For the HS model, the comoving Compton wavelength takes the form

\begin{equation}
\frac{\lambda_C}{1+z} = \sqrt{3(n+1)|f_{R0}|\frac{\bar{R}_0^{n+1}}{R^{n+2}}},
\end{equation}

\noindent and for a flat $\Lambda$CDM background its value today becomes 

\begin{equation}
\lambda_{C0} \approx 29.9\sqrt{\frac{|f_{R0}|}{10^{-4}}\frac{n+1}{4-3\Omega_m}} \,\, h^{-1}\text{Mpc},
\end{equation}

\noindent where $\Omega_m$ denotes the mean density of matter today in units of the critical density. For larger values of $n$ and a fixed $f_{R0}$, the Compton wavelength shrinks more
rapidly when going from $z=0$ to higher redshifts reducing the amount of
time for the modified forces to act on a given scale, and hence 
suppressing the enhanced growth compared to smaller $n$.  
For this reason, we expect that for larger $n$, larger $f_{R0}$ will be allowed by the data.

\subsection{Designer model}

\indent Another widely investigated class of $f(R)$ models are the
designer models, for which the functional form results from imposing a
specific expansion history (see e.g.\ \cite{2008PhRvD..77b3503P}). In
this work we restrict ourselves to spatially flat $\Lambda$CDM backgrounds. This family of models is commonly parametrized by the dimensionless Compton wavelength squared in Hubble units 

\begin{equation}\label{designer_compton}
B_0 \equiv \frac{f_{RR}}{1+f_R} R^{\prime} \frac{H}{H^{\prime}} \Big|_{z=0} \approx 2.1 \Omega_m^{-0.76}|f_{R0}|,
\end{equation}

\noindent with $f_{RR} = \text{d}f_R/\text{d}R$ and $^\prime \equiv \text{d}/\text{d}\ln a$. \\
\indent Despite the fact that both this and the previous class of
models reproduce either exactly or approximately the $\Lambda$CDM
background, their respective scalaron fields follow different
evolutions in time (see
e.g.\ \cite{2011PhRvD..83f3503F,2014AnP...526..259L}), and slightly
dissimilar modifications of gravity are provided by the two
cases. Therefore, one must be careful to compare only constraints from
the same class
(cf. \cite{2011PhRvD..83f3503F,2009PhRvD..80h3505S,2012PhRvD..85l4038L,2010JCAP...12..006A}). For
$f_{R0} \rightarrow 0$ and $B_0 \rightarrow 0$, both models reduce to $\Lambda$CDM, both in terms of expansion and growth. \\

\section{Mass function \label{sec:hmf}}

A self-consistent and accurate modeling of the mass function of dark
matter halos in terms
of the $f(R)$ parameters, $f_{R0}$ and $n$ or $B_0$, as well as the other
cosmological parameters is crucial to obtain proper constraints on
these parameters. The gold standard for predicting halo mass functions are N-body
simulations, which provide the reference values to which semi-analytical
predictions \cite{2008ApJ...688..709T,1999MNRAS.308..119S} are matched.  A breakthrough
occurred with the first consistent numerical simulations of $f(R)$ 
gravity \cite{2008PhRvD..78l3523O}, which have since been followed up with larger
and much higher resolution simulations \cite{2011PhRvD..83d4007Z,2012JCAP...01..051L,2013MNRAS.436..348P}.  Unfortunately, these
simulations are still very time consuming, and it is not feasible to
sample the cosmological parameter space using full simulations.  For this
reason, it is crucial to resort to physically motivated semi-analytical
approaches for the mass function predictions.  
\citet{2009PhRvD..79h3518S} presented a simple approach
based on both the spherical collapse approximation and the ST prescription, 
which they found to provide a good match to the mass function \emph{enhancement}
in $f(R)$ gravity relative to $\Lambda$CDM.  We will adopt this approach,
described in more detail below, to set conservative constraints on $f(R)$ gravity.\\
\indent The ST description for the comoving number density of halos per logarithmic interval of the virial mass $M_v$ is given by

\begin{equation}
n_{\Delta_v} \equiv \frac{dn}{d\ln M_v} = \frac{\bar{\rho}_m}{M_v}\frac{d\ln \nu}{d\ln M_v}\nu f(\nu).
\end{equation}

\noindent $\nu = \delta_c/\sigma(M_v)$ and $\delta_c$ are, respectively, the peak height and density thresholds, and

\begin{equation}
\nu f(\nu) = A\sqrt{\frac{2}{\pi}a\nu^2}\left[ 1 + (a\nu^2)^{-p} \right] \exp \left[ -a\nu^2/2 \right].
\end{equation}

\noindent $\sigma(M)$ is the variance of the linear matter density
field convolved with a top hat window function of radius $r$ that
encloses a mass $M = 4\pi r^3 \bar{\rho}_m/3$ for a given mean
background density $\bar{\rho}_m$,

\begin{equation}
\sigma^2(R,z) = \int \frac{d^3 k}{(2\pi^3)} |\tilde{W}(kr)|^2 P_L(k,z),
\end{equation}

\noindent where $P_L(k,z)$ is the linear power spectrum evolved to
redshift $z$ and $\tilde{W}(kr)$ is the Fourier transform of the window
function. The normalization constant is chosen such that $\int d\nu
f(\nu) = 1$. For $\Lambda$CDM, values of the ST mass function
parameters of $p = 0.3$, $a=0.75$, and $\delta_c = 1.673$
(corresponding to $\Omega_m=0.24$) have previously been shown to match
simulations at the $10$--$20\%$ level \cite{2009PhRvD..79h3518S}. The virial mass is defined as
the mass enclosed at the virial radius $r_v$, such that the average enclosed
density is $\Delta_{v}$ times the critical density of the Universe,
$\rho_c$. Equivalently, it is possible to use $\bar{\rho}_m$ rather than $\rho_c$
as a reference value, with the corresponding transformation between
both cases given by $\bar{\Delta}_{v} = \Delta_{v}/\Omega_m(z)$. The
virial mass can then be mapped into any other overdensity $\Delta$
assuming a Navarro-Frenk-White (NFW) halo mass profile with virial concentration $c
_v$ and using the
procedure outlined in \cite{2003ApJ...584..702H}. As shown in
\cite{2009PhRvD..79h3518S,2012PhRvD..85l4054L,2011PhRvD..83d4007Z}, within $r_v$ the
profiles of dark matter halos in $f(R)$ do not present any significant
deviation from those found in GR simulations, and therefore here we
can neglect $f(R)$ effects in the mass rescaling. In addition, the
exact value of the mass concentration has a negliglible effect on our
results as long as $c_{200} \gtrsim 3$. For this work we fix
$c_{200}=4$, as appropriate for the mass range of our data (see \cite{Mantz21012015} for more details). \\
\indent Our mass function calculation follows the approach adopted in \cite{2013JCAP...08..004S}. Deviations from GR are contained in a pre-factor given by the ratio of the ST mass function in $f(R)$ to that in GR

\begin{equation}\label{mfcn_tot}
n_\Delta = \left(  \frac{n_\Delta^{f(R)}}{n_\Delta^{\text{GR}}} \Bigg|_{\text{ST}} \right) n_{\Delta}|_{\text{Tinker}},
\end{equation}
 
\noindent with 

\begin{equation}
n_{\Delta}|_{\text{Tinker}} = \frac{\bar{\rho}_m}{M}\frac{d\ln\sigma^{-1}}{d\ln M} f(\sigma,z),
\end{equation}

\noindent and $f(\sigma,z)$ being the parametrization proposed and
fitted to GR simulations by \citet{2008ApJ...688..709T}. The latter includes the explicit redshift
dependence of the parameters and the covariance between them, as implemented
in \citet{2010MNRAS.406.1759M,Mantz21012015}, accounting for systematic uncertainties (such as the effects of
baryons\footnote{Using hydrodynamical simulations \citet{2014MNRAS.440..833A} showed that there is a bias between masses obtained using dynamical \mbox{methods} and those from lensing techniques, confirming the predictions of \citet{2010PhRvD..81j3002S}. As described in the main text, we account for this effect by calibrating our X-ray mass estimates with weak lensing data. In addition, the pre-factor in Eq.~\ref{mfcn_tot} could also be sensitive to the inclusion of baryonic physics into the calculation of the $f(R)$ HMF, for which only dark matter (DM) predictions currently exist.  \citet{2013MNRAS.436..348P}, however, estimated the impact of baryons on the matter power spectrum using hydrodynamical simulations. From their results one can show that, for scales $k \lesssim 10 \, h/\text{Mpc}$, $P_{\text{DM+baryons}}^{f(R)}/P_{\text{DM+baryons}}^{GR} \approx P_{\text{DM}}^{f(R)}/P_{\text{DM}}^{GR}$ demonstrating that the effects of baryons are similar for $f(R)$ and GR, and therefore negligible for their ratio (see also \cite{2015MNRAS.449.3635H}). The pre-factor of Eq.~\ref{mfcn_tot} should thus not be significantly affected by the presence of baryons.}, non-universality, etc.). Also, as explained in \cite{2008ApJ...688..709T}, the evolution in redshift of the mass function parameters is increasingly relevant for large overdensities (smaller radii). To attenuate this effect, we choose to work at a relatively large radius by setting $\Delta = 300\Omega_m(z)$. In Eq.~(\ref{mfcn_tot}), both the linear
variance, $\sigma(M)$, and the spherical collapse parameters are
calculated using the corresponding theory of gravity, either GR or $f(R)$. For $\delta_c$, we adopt the following fitting formula \cite{1997PThPh..97...49N} 

\begin{equation}
\delta_c(\Omega_m,z) =  \mathcal{A}\left(  1-\mathcal{B}\log_{10}\left[ 1+\frac{\Omega_m^{-1}-1}{(1+z)^3} \right] \right),
\end{equation}

\noindent with $\mathcal{A}=1.6865$ and $\mathcal{B} = 0.0123$ for
GR, and $\mathcal{A}=1.7063$ and $\mathcal{B}=0.0136$ for $f(R)$. The latter values were calculated assuming a spherical
perturbation smaller than the local Compton
wavelength and forces enhanced by $4/3$ everywhere and for all
epochs, and therefore are independent of the particular choice of $f(R)$ model. Using N-body simulations, \citet{2009PhRvD..79h3518S} showed that in the large-field regime ($|f_{R0}| \gtrsim 10^{-5}$) these values provide an underestimate of the effect on the mass function, and will thus yield conservative upper limits on
$|f_{R0}|.$\footnote{Even though this HMF was originally calibrated for the HS model with $n=1$, \citet{2011PhRvD..83f3503F} showed that for the regimes of interest here, large-field (linear) and transition, this HMF can also be safely used for other values of $n$, and by extension for the designer model by correspondingly adjusting only the linear term $\sigma(M,z)$. The results on the matter power spectrum for the HS and designer models from \citet{2013PhRvD..88j3507H} give also additional support to the latter conclusion.} In addition, in order to model the GR limit we set $n^{f(R)}_{\text{ST}}/n^{\text{GR}}_{\text{ST}}$ to 1 whenever this ratio becomes smaller than 1. Effectively, this approximation introduces a screening mechanism that is much more efficient than the one predicted by simulations, allowing larger values of $f_{R0}$ to be consistent with the data. A
less conservative approach would be to model the chameleon
mechanism, which would change the predictions for the mass function when
$|f_{R0}| \lesssim 10^{-5}$. Note, however, that entering this regime without properly validating the modeling of the chameleon suppression with simulations might result in spuriously tight constraints. We leave the accurate modeling of the mass function in this regime for future work (Cataneo et al., in preparation). This will then allow us to explore the rest of the parameter space currently available to clusters, and to cosmological data by extension. See also \cite{2014JCAP...03..021L,2011PhRvD..84h4033L} for recent approaches to modeling the chameleon mechanism. \\
\indent Lastly, note that, to calculate $\Delta_v^{f(R)}$, we use the fitting formula valid for flat $\Lambda$CDM \cite{1998ApJ...495...80B}

\begin{equation}
\Delta_{v}^{\text{GR}}(\Omega_{mz}) = 18\pi^2 - 82(1-\Omega_{mz}) - 39(1-\Omega_{mz})^2,
\end{equation}

\noindent with $\Omega_{mz} \equiv \Omega_m(z)$, and fix the ratio
$\Delta_v^{f(R)}/\Delta_v^{\text{GR}}$ to 74/94 \cite{2009PhRvD..79h3518S}. We have checked that
this scaling is a good approximation (better than 2 per cent) for a range of $0.1 < \Omega_{m} < 0.6$, which is much wider than the constraints on this quantity set by our cluster data alone (see \cite{Mantz21012015}), and for a redshift range of $0 < z < 0.7$, which extends beyond that of our cluster growth data.

\section{Data \label{sec:data}}

\begin{figure*}[t]
\includegraphics[width=0.9\columnwidth]{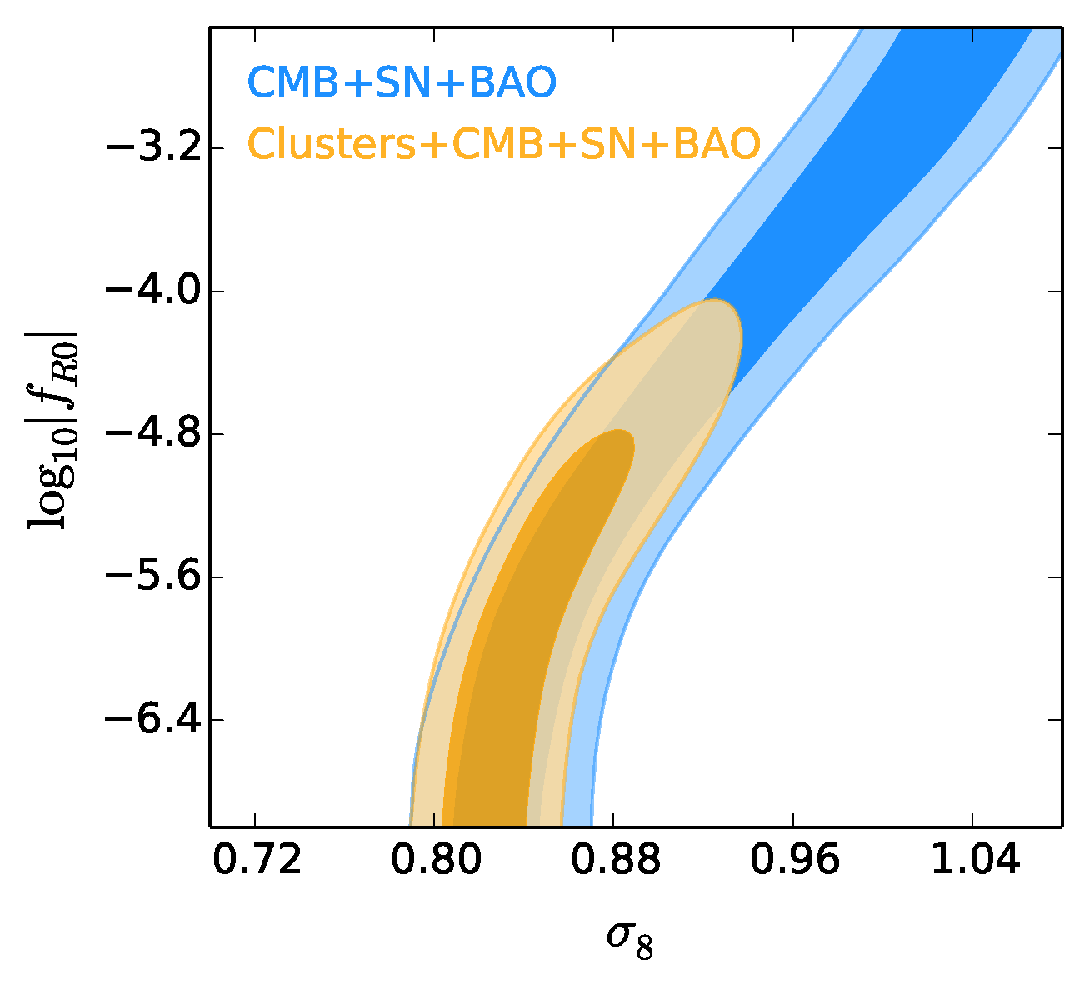}
\qquad
\includegraphics[width=0.9\columnwidth]{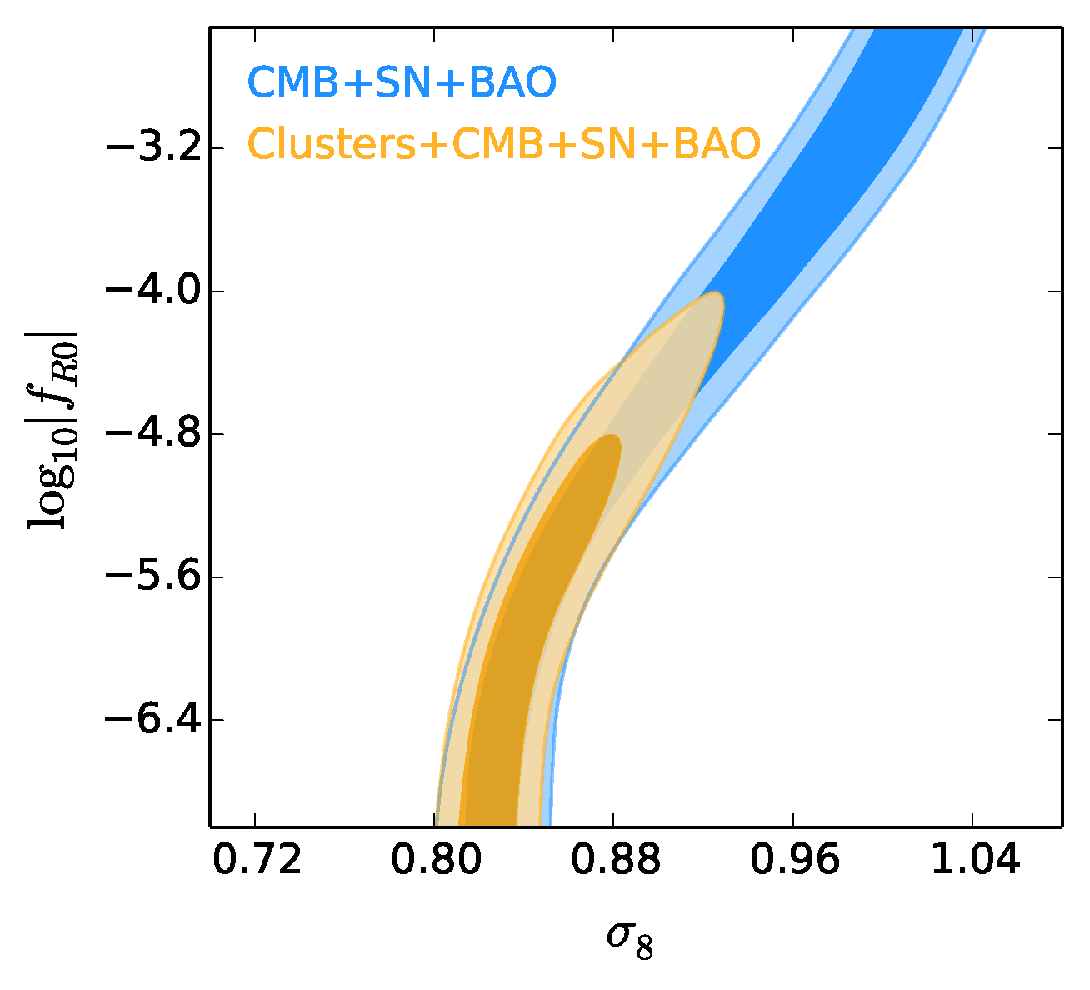}
\caption{Constraints on the HS model with $n=1$. Dark and light shadings indicate the 68.3 and 95.4 per cent confidence regions (accounting for systematic uncertainties) from the following data sets: the CMB combined with SNIa+BAO (blue), and the combination of all these with clusters (gold). In the left panel, we use WMAP+ACT+SPT as CMB data, and \textit{Planck}+WP+lensing+ACT+SPT in the right panel.}
\label{HS1_cl_cmb_comb}
\end{figure*}

\subsection{Cluster data}

For the cluster growth analysis we employ the \textit{ROSAT} Brightest Cluster Sample [BCS; $z<0.3$ and $F_X(0.1$--$2.4\, \text{keV}) > 5\times 10^{-12}$ \text{erg\,s$^{-1}$\,cm$^{-2}$}] \cite{1998MNRAS.301..881E}, the \textit{ROSAT}-ESO Flux Limited X-ray sample [REFLEX; $z<0.3$ and $F_X(0.1$--$2.4\, \text{keV}) > 3\times 10^{-12}$ \text{erg\,s$^{-1}$\,cm$^{-2}$}] \cite{2004A&A...425..367B}, and the Bright sample of the Massive Cluster Survey [Bright MACS; $0.3 < z < 0.5$ and $F_X(0.1$--$2.4\, \text{keV}) > 2\times 10^{-12}$ \text{erg\,s$^{-1}$\,cm$^{-2}$}] \cite{2010MNRAS.407...83E}. In order to reduce systematic uncertainties, a few detections later found to have their X-ray emission dominated by point sources (active galactic nuclei) rather than the intracluster medium have been removed, and higher flux limits have been applied to avoid incompleteness when selecting clusters from BCS (cf. \cite{2010MNRAS.406.1759M,Mantz21012015}). Overall, the sample contains a total of 224 clusters. For 94 of these clusters X-ray luminosities and gas masses from \textit{ROSAT} and/or \textit{Chandra} data (see \cite{2010MNRAS.406.1773M} for details) were used to constrain cluster scaling relations and take full advantage of the mass information available for individual clusters \cite{Mantz21012015}. \\
\indent For the calculation of the absolute cluster mass scale we use state-of-the-art weak gravitational lensing measurements for 50 massive clusters (see
\cite{Mantz21012015,2014MNRAS.439....2V,2014MNRAS.439...28K,2014MNRAS.439...48A}
for details). As discussed above, since for the relevant field regime the lensing mass in $f(R)$ is the same as in
GR up to currently undetectable effects of order $f_{R0}$, we do not need to apply any correction on the mass function due to
the effect of the fifth force on the mass estimates \cite{2010PhRvD..81j3002S}. \\
\indent We also employ X-ray measurements of the gas mass fraction,
$f_{\text{gas}}$, in a shell of 0.8 to 1.2 times the radius corresponding to a critical overdensity $\Delta=2500$ for a sample of the hottest, most X-ray
luminous and dynamically relaxed galaxy clusters
\cite{2014MNRAS.440.2077M}. These data add constraining power on the
background expansion model, and on
$\Omega_m$, which helps break the degeneracy of the normalization of the matter power spectrum $\sigma_8 \equiv \sigma(r=8 h^{-1}\text{Mpc},z=0)$ with this parameter. In this experiment, cluster masses are also calibrated using weak lensing data, in order to constrain instrumental (calibration) and astrophysical (bias due to the assumption of hydrostatic equilibrium) systematics. \\ 
\indent As shown in  \cite{2010PhRvD..81j3002S}, we could also employ our measurements of the ratio between lensing and X-ray mass estimates to constrain $f_{R0}$. In our current analysis, this signal would be completely degenerate with our instrumental and astrophysical uncertainties, and from our present estimates of these systematics, we would have little constraining power on $f_{R0}$. However, this is a promising new avenue for the near future.
\subsection{CMB data}

For the analyses including CMB data, we use measurements from either the
\textit{Wilkinson Microwave Anisotropy Probe} (WMAP 9-year release;
\cite{2013ApJS..208...20B,2013ApJS..208...19H}) or the \textit{Planck}
satellite (year-1 release plus WMAP polarization data, hereafter
denoted as \textit{Planck}+WP; \cite{2014A&A...571A..15P}). We also use data from the gravitational lensing
potential generated by large scale structures, as measured by the
\textit{Planck} Collaboration \cite{2014A&A...571A..17P}. We refer to the combination of these with \textit{Planck}+WP power spectrum data as \textit{Planck}+WP+lensing. Our two complete sets of CMB data also include high multipole measurements from the Acatama Cosmology Telescope (ACT; \cite{2014JCAP...04..014D}) and the South Pole Telescope (SPT; \cite{2011ApJ...743...28K,2012ApJ...755...70R,2013ApJ...779...86S}). \\
\indent When using CMB data, we also fit for the cosmic baryon and
dark matter densities, $\Omega_b h^2$ and $\Omega_c h^2$; the optical
depth to reionization, $\tau$; the amplitude and spectral index of the
scalar density perturbations, $A_s$ and $n_s$; and the characteristic
angular scale of the acoustic peaks, $\theta$  (which effectively determines $H_0$). We also marginalize over the set of nuisance parameters
associated with each CMB data set, accounting for the thermal Sunyaev-Zel'dovich effect and unresolved foregrounds.

\subsection{Additional data sets}

Certain parameter degeneracies relevant at late times, like the one
between $f_{R0}$ and $\Omega_m$, can be helped by including additional
cosmological distance probes, such as those using SNIa and BAO data. We use
the Union 2.1 compilation of SNIa \cite{2012ApJ...746...85S}, and BAO
data from a combination of measurements from the 6-degree Field Galaxy Survey (6dF; $z=0.106$; \cite{2011MNRAS.416.3017B}), the Sloan Digital Sky Survey (SDSS; $z=0.35$ and $z=0.57$; \cite{2012MNRAS.427.2132P,2014MNRAS.439...83A}), and the WiggleZ Dark Energy Survey ($z=0.44,0.6$ and 0.73; \cite{2011MNRAS.418.1707B}). Note, however, that including these additional data sets affects our results only when we use WMAP+ACT+SPT as a CMB data set. In this case, we find that the addition of SNIa+BAO data helps in breaking the degeneracy with $\Omega_m$ and improves our constraints on $f_{R0}$ or $B_0$. If instead of WMAP we use \textit{Planck}+WP, the impact of adding SNIa+BAO data is negligible (see section~\ref{sec:results}).

\section{Results \label{sec:results}}

We obtain the posterior probability distribution functions (pdf) of our parameters using the MCMC engine {\footnotesize \textrm{COSMOMC}}\footnote{\url{http://cosmologist.info/cosmomc/}} \cite{2002PhRvD..66j3511L} (October 2013 version), but modified to include two additional likelihood modules, one for $f_{\text{gas}}$ data\footnote{\url{http://www.slac.stanford.edu/~amantz/work/fgas14/}} and the other for cluster growth data \cite{Mantz21012015}. Hereafter we will refer both of them together as cluster data. To calculate the evolution of the cosmic mean background density and its linear perturbations we use {\footnotesize \textrm{MGCAMB}}\footnote{\url{http://www.sfu.ca/~aha25/MGCAMB.html}} \cite{2009PhRvD..79h3513Z,2011JCAP...08..005H}, which is an extension of the Boltzmann code {\footnotesize \textrm{CAMB}}\footnote{\url{http://camb.info}} \cite{2000ApJ...538..473L} that includes modified gravity models. We have also implemented the HS model\footnote{\url{http://icosmology.info/HuSawicki.html}} into  {\footnotesize \textrm{MGCAMB}}, and a few corresponding modifications to facilitate the calculations of secondary anisotropies of the CMB generated by the modified growth of structure.\\
\indent Throughout our analysis, we assume the minimal value of the species-summed neutrino mass allowed by neutrino oscillation measurements in the normal hierarchy, $\sum m_\nu = 0.056$ eV, and the standard effective number of relativistic species, $N_{\text{eff}} = 3.046$. Massive neutrinos suppress structure formation on scales smaller than the free streaming scale, and this effect can counteract the enhancement introduced by $f(R)$ modifications of gravity, allowing larger $f_{R0}$ values currently excluded \cite{2014MNRAS.440...75B, 2013PhRvL.110l1302M}. In order to use cluster data to test $f(R)$ models while also allowing $\sum m_{\nu}$ and $N_{\text{eff}}$ to be free parameters would require an accurate HMF validated by simulations that incorporates simultaneously both extensions of $\Lambda$CDM. Note, though, that the minimal neutrino mass adopted in the present work is too small to significantly alter our HMF.\\
\indent For the present-day amplitudes of the scalaron field in each modified gravity model, we employ the following uniform priors: $\log_{10}B_0 \in [-10,0.5]$ and $\log_{10}|f_{R0}| \in [-10,-2.523]$. Since from theory we have no information on the order of magnitude of the modification (see also \cite{2014arXiv1409.6530C,2014JCAP...03..046D}), we use logarithmic priors, which weight all scales equally. Note, however, that GR ($B_0=0$ or $f_{R0}=0$) is in practice unreachable in log space, and therefore the results for $\log_{10}B_0$ or $\log_{10}|f_{R0}|$ will be dependent on the lower bound of the prior. Using the combination \textit{Planck}+WP+lensing+SNIa+BAO, for the ``designer'' model we have explicitly checked the dependence of the marginalized pdf on the lower bound of the log-prior for two different values, $[-10,0.5]$ and $[-7,0.5]$. The resulting upper limits on $\log_{10} B_0$ show a difference of about 10 per cent. We have also run MCMC chains with uniform priors on $B_0$ showing that, as expected, in these cases we obtain upper limits that are about an order of magnitude larger than those for the log-priors\footnote{Intuitively,  this can be understood by applying a change of variable to convert the linear to the logarithmic pdf (or vice versa). Going from $f_{R0}$ ($B_0$) to $\log_{10}|f_{R0}|$ ($\log_{10}B_0$) exponentially suppresses the probability for small parameter values due to the Jacobian of the transformation. If one uses directly a log-prior all scales will contribute to the pdf correspondingly lowering the upper limit. One can also directly convert the MCMC scalaron amplitude values from linear to log, accounting for the Jacobian of the transformation. The pdf obtained from the resulting chains will be approximately equivalent to that calculated from chains using a log-prior with a lower bound determined by matching the two pdfs. Note that this bound will be related to the tail of the linear run, which is characterized by the constraining power of the data.}. It is therefore important to fully state the priors used in the analysis in order to allow others to properly compare results.

\begin{figure}[t]
\includegraphics[width=1.0\columnwidth]{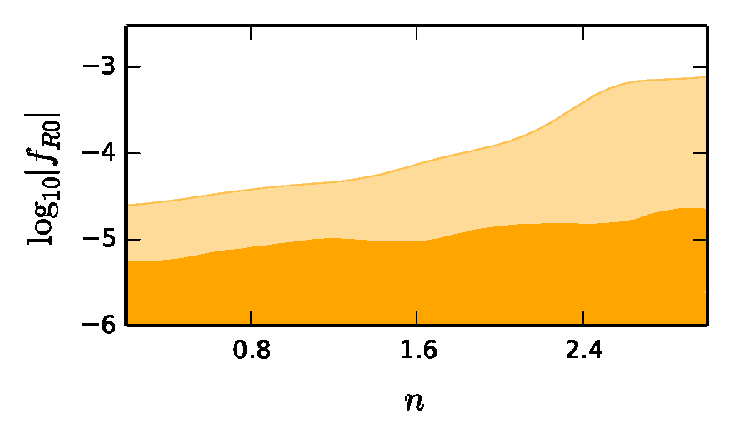}
\caption{Constraints on the HS model with varying $n$. Dark and light shadings indicate the 68.3 and 95.4 per cent confidence regions (accounting for systematic uncertainties) from the combination of clusters, CMB (\textit{Planck}+WP+lensing+ACT+SPT) and SNIa+BAO.}
\label{HSnvar_cl_pl_lens_actspt_comb}
\end{figure}

\begin{figure*}[t]
\includegraphics[width=0.9\columnwidth]{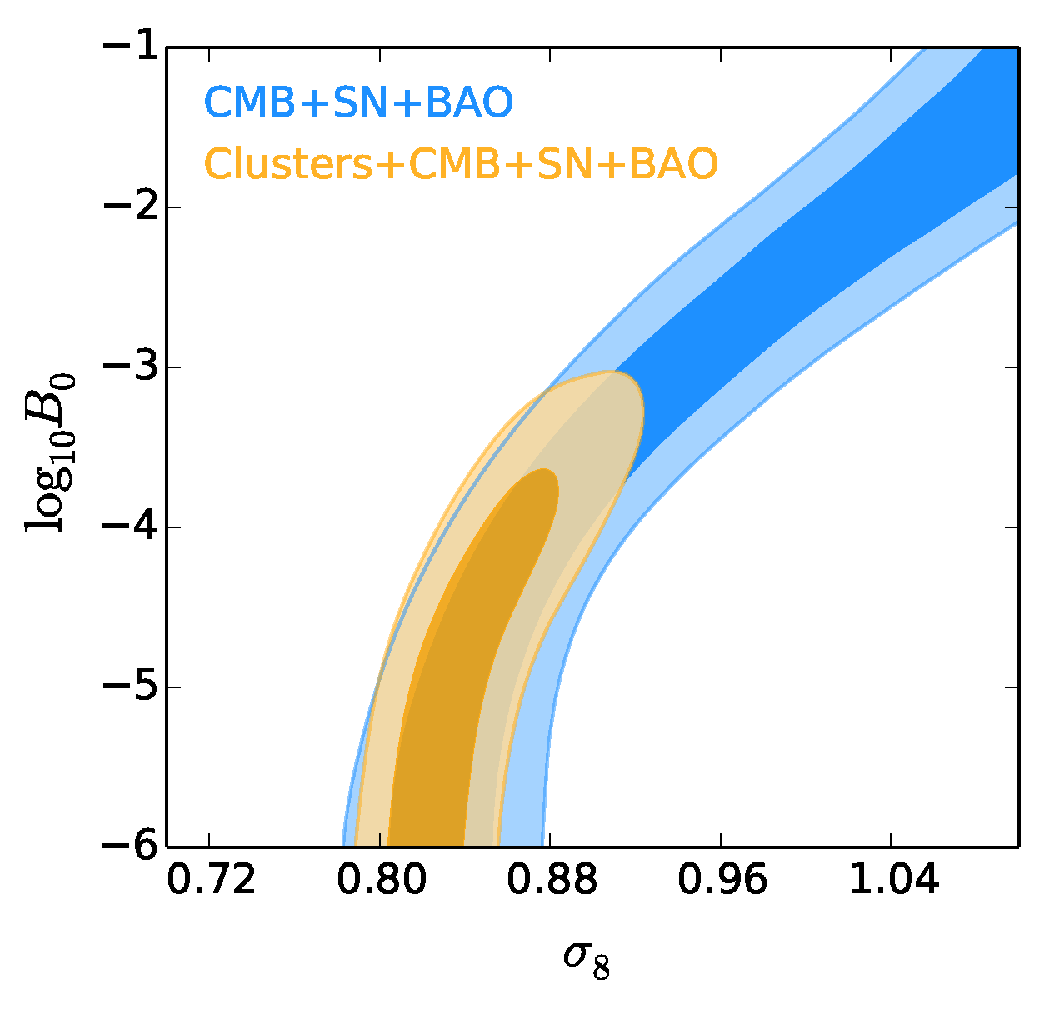}
\qquad
\includegraphics[width=0.9\columnwidth]{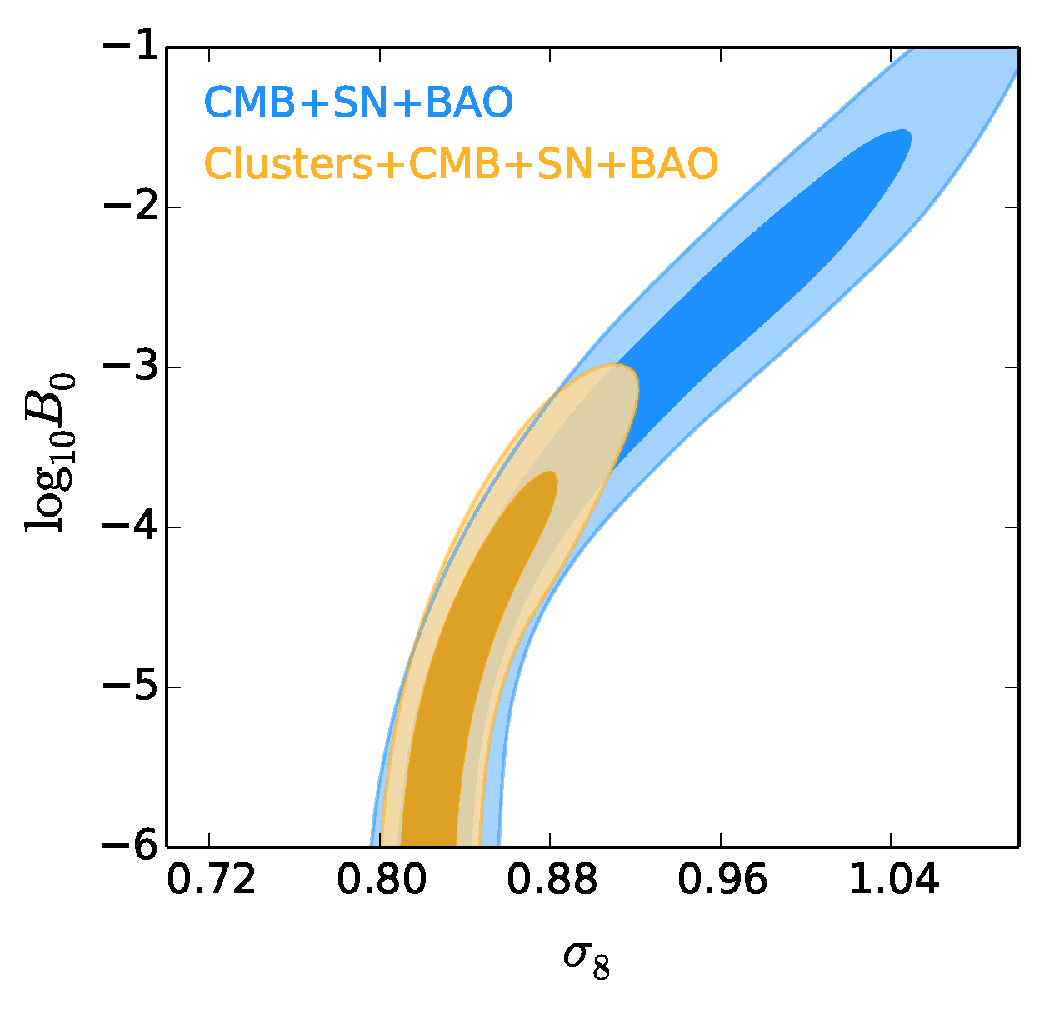}
\caption{Constraints on the designer model. Dark and light shadings indicate the 68.3 and 95.4 per cent confidence regions (accounting for systematic uncertainties) from the following data sets: clusters (purple), the CMB plus SNIa+BAO (blue), and the combination of all these (gold). In the left panel, we use WMAP+ACT+SPT as CMB data, and \textit{Planck}+WP+lensing+ACT+SPT in the right panel.}
\label{des_cl_pl_lens_actspt_comb}
\end{figure*}

\indent For the HS model with $n=1$, fig.~\ref{HS1_cl_cmb_comb} shows the joint constraints on $f_{R0}$ and $\sigma_8$ from the CMB (blue contours; including also SNIa and BAO) and from these plus clusters (gold contours). For large values of $f_{R0}$, CMB data present a clear degeneracy between $f_{R0}$ and $\sigma_8$. For $|f_{R0}| \lesssim 10^{-6}$ we recover as expected the same values of $\sigma_8$ as those obtained for GR. This is because in this regime the variance of the linear matter fluctuations on a scale of $8 h^{-1}\text{Mpc}$ becomes insensitive to the modifications of gravity.\\
\indent Given the use of clusters and the CMB, the addition of SNIa and BAO data impacts on our results mainly by constraining $\Omega_m$. When we use clusters plus \textit{Planck}+WP+lensing+ACT+SPT, the impact of including SNIa+BAO data is negligible since the combined $\Omega_m$ constraints are essentially unchanged. However, for the combination of clusters with WMAP+ACT+SPT, the inclusion of SNIa+BAO data sets shifts the constraints on $\Omega_m$ to higher values providing similar results to those obtained from the combination with \textit{Planck} data.\\
\indent Cluster data provides strong measurements on the growth of structure at late times when the modifications of gravity are relevant. The main contribution of the CMB to the combined results is to tighten the constraints on matter power spectrum parameters such as $A_s$ and $\Omega_m$, which consequently allow clusters to break the degeneracy between $f_{R0}$ and $\sigma_8$ by constraining the latter, and thus providing a tight upper limit on the scalaron amplitude. This is clear in figs.~\ref{HS1_cl_cmb_comb} and~\ref{des_cl_pl_lens_actspt_comb} by comparing the constraints without and with clusters (blue and gold contours, respectively). Using WMAP+ACT+SPT as the CMB data set, we obtain $\log_{10}|f_{R0}|<-4.73$, and using \textit{Planck}+WP+lensing+ACT+SPT we have $\log_{10}|f_{R0}|<-4.79$ (see also Table~\ref{table1}).

\indent The CMB constraints on the left panel of fig.~\ref{HS1_cl_cmb_comb} correspond to WMAP+ACT+SPT data, and those on the right panel to \textit{Planck}+WP+lensing+ACT+SPT data. The higher precision of the measurements from \textit{Planck} improves the constraints on many of the non-gravity specific cosmological parameters and ultimately on $\sigma_8$, as shown by comparing these two panels. As pointed out in \cite{2014JCAP...03..046D}, without the lensing potential data, large $f_{R0}$ values are preferred due to lower power in the low multipoles and higher lensing signal in the high multipoles of the \textit{Planck} temperature power spectrum. The addition of the lensing potential data, which probes scales in the range $10^{-2}<k<10^{-1} \, h/\text{Mpc}$ at $z \sim 2$, disfavors large values of $f_{R0}$, while keeping the constraints on the other cosmological parameters essentially unchanged.\\
\indent We have also run a more general analysis for the HS model including $n$ as an additional free parameter with a uniform prior of $ 0.2 \leq n \leq 3 $. As expected and shown in fig.~\ref{HSnvar_cl_pl_lens_actspt_comb}, for increasing $n$ the constraints on $f_{R0}$ become weaker due to a growth of structure that is asymptotically closer to GR.  Nonetheless, our results indicate a greater constraining power from the current data than the conservative projections in \cite{2011PhRvD..83f3503F}.  \\
\indent For the designer model we find similar results. Fig.~\ref{des_cl_pl_lens_actspt_comb} shows that the combination of cluster and CMB data, either from WMAP+ACT+SPT (left panel) or from \textit{Planck}+WP+lensing+ACT+SPT (right panel), constrains the background Compton wavelength to a few tens of megaparsecs ($\log_{10}B_0<-3.75$ and $\log_{10}B_0<-3.68$, respectively). As shown before \cite{2014JCAP...03..046D}, we also find that adding the CMB lensing potential data to the combination of \textit{Planck}+WP+ACT+SPT places a mild upper limit on $B_0$ (see e.g. the right panel of fig.~\ref{des_cl_pl_lens_actspt_comb}). However, for the HS model the same data combination does not provide an upper limit on $|f_{R0}|$ at the value that one would expect from naively using eq. 6 to convert the limit obtained on $B_0$ for the designer model. This is due to the different evolution of the Compton wavelength in the two models.\\
\indent Table \ref{table1} summarizes the upper limits on $f_{R0}$ and $B_0$\footnote{Because their growth histories are similar, although not identical, note that the constraints on HS models with n=1 and designer models are comparable. An approximate conversion between $f_{R0}$ and $B_0$ can be achieved using Eq. \ref{designer_compton}.} for the combinations of data sets used in this work, which are compatible with those obtained combining CMB and matter power spectrum measurements \cite{2014JCAP...03..046D, PhysRevD.91.063524}. These limits are arguably the most robust to date using the abundance of galaxy clusters and unlike previous work \cite{2009PhRvD..80h3505S,2012PhRvD..85l4038L} push the constraints into the transition regime where the most massive halos are screened.\\

\begin{table*}[t]
\caption{Marginalized 95.4 per cent upper limits on $f(R)$ parameters for the two models discussed in the text, Hu-Sawicki (HS) and designer.}
\label{table1}
\begin{center}
\renewcommand{\arraystretch}{1.3}
\begin{ruledtabular}
\begin{tabular}{c >{\centering\arraybackslash}p{0.08\textwidth} >{\centering\arraybackslash}p{0.1\textwidth} >{\centering\arraybackslash}p{0.2\textwidth}}
				Data 							& 			\multicolumn{2}{ c }{HS model}						&		Designer model 			\\ 
												& 		$\log_{10}|f_{R0}|$ 		& 		    $n$			&		$\log_{10}B_0$ 	\\  \hline
	Clusters+WMAP+ACT+SPT+SNIa+BAO						&			-4.73			&		      1				&			-3.75	\\
	Clusters+\textit{Planck}+WP+lensing+ACT+SPT+SNIa+BAO 	&			-4.79			&		      1				&			-3.68			\\			
	Clusters+\textit{Planck}+WP+lensing+ACT+SPT+SNIa+BAO 	&			-3.95				& 	$0.2 \leq n \leq 3$ 		&						\\		
\end{tabular}
\end{ruledtabular}
\renewcommand{\arraystretch}{1}
\end{center}
\label{default}
\end{table*}

\section{Conclusions \label{sec:conclusions}}

\indent We have performed a full, self-consistent joint MCMC likelihood analysis for two $f(R)$ models, Hu-Sawicki (HS) and ``designer". These two models mimic either closely or exactly the expansion history of $\Lambda$CDM, but deviate with respect to its growth history. Our results are driven by the combination of galaxy cluster and CMB data, to which we also add other data sets. The abundance of massive galaxy clusters is a powerful cosmological probe of gravity on scales that are inaccessible to local and astrophysical tests of gravity, and its sensitivity derives from the steepness of the high mass tail of the halo mass function. The CMB data provide tight measurements on the matter power spectrum at high redshifts that together with those from the cluster data at low redshifts allow us to break key degeneracies and constrain $f(R)$ modifications on the growth rate at late times.\\
\indent In the context of $f(R)$ gravity, departures from GR are sourced by an additional scalar degree of freedom responsible for an effective fifth force that enhances the growth of structures for scales smaller than its Compton wavelength. As a result, the abundance of massive halos increases for amplitudes of the background scalar field $|f_{R0}| \gtrsim 10^{-6}$; below this value, the chameleon screening mechanism leads to a negligible modification of the abundance of massive clusters.\\
\indent We use constraints on the expansion and growth histories from cluster abundance data, and on the expansion history from $f_{\text{gas}}$ data. For the latter, it is interesting to note that a comparison between the dynamical masses derived from X-ray data and the weak lensing mass calibration \cite{2010PhRvD..81j3002S} could also be included in the $f(R)$ analysis to add constraining power in the large-field regime, and to possibly help breaking parameter degeneracies. In particular, while massive neutrinos can partially counteract the effects of $f(R)$ gravity on the abundance of galaxy clusters, these will not lead to a mismatch between their lensing and X-ray masses. This promising measurement is currently limited by instrumental and astrophysical uncertainties in the determination of our X-ray masses. In order to make this option viable, we will therefore need to reduce these systematic uncertainties by e.g. using new X-ray line emission data from the upcoming \textit{Astro-H} mission to measure residual bulk motions. Additional lensing data will then ensure us sufficient constraining power on $f(R)$. \\
\indent From the combination of cluster and CMB data, either from \textit{Planck}+WP (or WMAP) plus ACT+SPT, and including also SNIa+BAO data, we obtain tight upper bounds $\log_{10}|f_{R0}| < -4.79$ (or -4.73) for the HS model (with $n=1$) and $\log_{10}|B_0| < -3.68$ (or -3.75) for the designer model. Our results are obtained using high quality cluster growth data up to $z\sim0.5$, a tight control of systematic uncertainties, a robust mass calibration from weak lensing data, and the full shape of the halo mass function for the mass range of our data. Including CMB data is essential to significantly tighten the constraints on cosmological parameters such as $A_s$ and $\Omega_m$, which then enables clusters to break a remaining key degeneracy between $\sigma_8$ and $f_{R0}$ ($B_0$). SNIa and BAO data are only relevant when WMAP+ACT+SPT is used as a CMB data set. In this case, the addition of the SNIa+BAO data provides similar constraints on $\Omega_m$, and consequently on $f_{R0}$, to those obtained with the combination that instead of WMAP has \textit{Planck} data.\\
\indent For the near future, further progress using current cluster data is within reach. Primarily, this will require an accurate modeling of the Chameleon screening mechanism in high density environments as a function of standard cosmological and model parameters, halo mass, and redshift. Testing the resulting theoretical prediction for the HMF against cosmological simulations for different cosmologies will be crucial to assess the accuracy of this result (Cataneo et al., in preparation).\\ 
\indent A self-consistent implementation of the non-linear Chameleon suppression of $f(R)$ into our cluster likelihood analysis should reduce the current upper limits by about another order of magnitude, below which data limited to relatively low redshift massive galaxy clusters cannot distinguish between GR and $f(R)$ gravity. \\
\indent Ongoing and planned surveys will also be able to improve further $f(R)$ constraints. The Dark Energy Survey \cite{2005astro.ph.10346T}, Euclid \cite{2011arXiv1110.3193L} and the Large Synoptic Survey Telescope \cite{2008arXiv0805.2366I} in the optical, the eROSITA all-sky survey \cite{2012arXiv1209.3114M} in the X-ray, and Sunyaev-Zel'dovich effect surveys (such those from \textit{Planck} \cite{2014A&A...571A..29P}, the South Pole Telescope \cite{2015ApJS..216...27B}, and the Atacama Cosmology Telescope \cite{2013JCAP...07..008H}) in the mm/submm will substantially expand both the mass and redshift range of cluster samples, including identifying the most massive clusters up to $z\sim2$. This will allow us to probe all the relevant evolution of the Compton wavelength and extend the measured mass function to masses where departures from GR are significant in the regime $|f_{R0}| \lesssim 10^{-6}$ due to the inefficiency of the chameleon screening mechanism.

\begin{acknowledgments}
MC thanks A. Agnello, N. C. Amorisco, M. Barnab\`e, C. Grillo and R. Wojtak for fruitful discussions on the effects of prior probability distributions. The computational analysis was performed using the High Performance Computing (HPC) facility at the University of Copenhagen, and the Gardar supercomputer of the Nordic HPC project. The Dark Cosmology Centre (DARK) is funded by the Danish National Research Foundation. ABM was supported by the National Science Foundation under grant  AST-1140019.
\end{acknowledgments}

\bibliographystyle{apsrev4-1}
\bibliography{reference_aps}

\end{document}